\PassOptionsToPackage{table,x11names}{xcolor}
\documentclass{l4dc2024}
\title[Learning Policies for upMDPs]{Learning Robust Policies for Uncertain Parametric \\Markov Decision Processes}
\usepackage{times}

\usepackage{times}

\newcommand\given[1][]{\:#1\vert\:}
\newcommand\PCTLp[1][]{\mathsf{P}}
\newcommand\until[1][]{\mathsf{U}}
\newcommand\PCTLnext[1][]{\mathsf{X}}
\newcommand\finally[1][]{\mathsf{F}}
\newcommand\states[1][]{\mathcal{S}}
\newcommand\acts[1][]{\mathcal{A}}
\newcommand\probs[1][]{T}
\newcommand\traj[1][]{\zeta}

\newcommand\paramProbs[1][]{\mathcal{T}}
\newcommand\MDP[1][]{\mathcal{M}}
\newcommand\upMDP[1][]{\mathcal{M}_v^\mathbb{P}}
\newcommand\sampledMDP[1][]{\mathcal{M}[v]}
\newcommand\params[1][]{\mathcal{V}}
\newcommand\probDist[1][]{\mathbb{P}}
\newcommand\allSamples[1][]{\mathcal{U}_N}
\newcommand\finiteSampledMDP[1][]{\mathcal{M}[u]}
\newcommand\optSampledMDP[1][]{\mathcal{M}[u^\star]}

\newcommand\sol[3][]{\text{sol}_{\MDP}^{#2}(#3;\pathForm)}

\newcommand\pol[1][]{\pi}
\newcommand\pols[1][]{\Pi}
\newcommand\stateForm[1][]{\Phi}
\newcommand\pathForm[1][]{\psi}
\newcommand\supps[1][]{\Tilde{s}^\ast_N}
\newcommand\AllSuppSet[1][]{\Sigma^\ast}
\newcommand\lbEps[1][]{\underline{\epsilon}(\supps)}
\newcommand\ubEps[1][]{\overline{\epsilon}(\supps)}
\newcommand\mixedProb[1][]{\pi^M(\pi^D)}
\newcommand\behaviouralProb[1][]{\pi^B(s)(a)}

\newtheorem{problem}{Problem}

\DeclareMathOperator*{\argmax}{arg\,max}
\DeclareMathOperator*{\argmin}{arg\,min}
\DeclareMathOperator*{\spacer}{\vert}
\DeclareMathOperator*{\sat}{\models}
\DeclareMathOperator*{\notsat}{\not\models}
\DeclareMathOperator*{\Dist}{\text{Dist}}

\usepackage{cleveref}
\usepackage{wrapfig}
\usepackage{multirow}

\crefname{algorithm}{Algorithm}{Algorithm}
\crefname{pluralalgorithm}{Algorithms}{Algorithms}

\crefname{algocf}{Algorithm}{Algorithm}

\crefname{figure}{Figure}{Figures}
\crefname{pluralfigure}{Figures}{Figures}

\crefname{section}{Section}{Sections}
\crefname{pluralsection}{Sections}{Sects}

\crefname{table}{Table}{Table}
\crefname{pluraltable}{Tables}{Tables}

\crefname{definition}{Definition}{Definition}
\crefname{pluraldefinition}{Definitions}{Definitions}

\crefname{theorem}{Theorem}{Theorems}
\crefname{pluraltheorem}{Theorems}{Theorems}

\crefname{lemma}{Lemma}{Lemmas}
\crefname{plurallemma}{Lemmas}{Lemmas}

\crefname{example}{Example}{Example}
\crefname{pluralexample}{Examples}{Examples}

\crefname{problem}{Problem}{Problem}
\crefname{pluralproblem}{Problems}{Problems}

\crefname{assumption}{Assumption}{Assumption}
\crefname{pluralassumption}{Assumptions}{Assumptions}

\crefname{assumptionx}{Assumption}{Assumption}
\crefname{pluralassumptionx}{Assumptions}{Assumptions}

\crefname{remark}{Remark}{Remark}
\crefname{pluralremark}{Remarks}{Remarks}

\crefname{appendix}{Appendix}{Appendices}
\crefname{pluralappendix}{Appendices}{Appendices}

\usepackage{tikz}
\usetikzlibrary{shapes.geometric, arrows.meta, positioning}
\tikzstyle{startstop} = [rectangle, rounded corners, minimum width=3cm, minimum height=0.1cm,text centered, draw=black, fill=pink, text width=4cm]
\tikzstyle{IO} = [ellipse, minimum width=2cm, minimum height=0.1cm,text centered, draw=black, text width=3cm]
\tikzstyle{arrow} = [thick,->,>={Stealth}]
\tikzset{font={\fontsize{10pt}{12}\selectfont}}

\usepackage{mathtools}

\usepackage{enumitem}
\setlist{topsep=0pt, leftmargin=*}

\usepackage{adjustbox}

\usepackage{array}
\newcolumntype{x}[1]{>{\centering\arraybackslash\hspace{0pt}}p{#1}}

\usepackage{caption}

\newif\ifappendix
  \global\appendixtrue

\coltauthor{\Name{Luke Rickard} \Email{luke.rickard@eng.ox.ac.uk}\\
 \Name{Alessandro Abate} \Email{alessandro.abate@cs.ox.ac.uk}\\
 \Name{Kostas Margellos} \Email{kostas.margellos@eng.ox.ac.uk}\\
 \addr Department of Engineering Science, University of Oxford, Oxford, UK}

\begin{document}

\maketitle
\begin{abstract}%
Synthesising verifiably correct controllers for dynamical systems is crucial for safety-critical problems. 
To achieve this, it is important to account for uncertainty in a robust manner, while at the same time it is often of interest to avoid being overly conservative with the view of achieving a better cost.
We propose a method for verifiably safe policy synthesis for a class of finite state models, under the presence of structural uncertainty. 
In particular, we consider uncertain parametric Markov decision processes (upMDPs), a special class of Markov decision processes, with parameterised transition functions, where such parameters are drawn from a (potentially) unknown distribution.
Our framework leverages recent advancements in the so-called scenario approach theory, where we represent the uncertainty by means of scenarios, and provide guarantees on synthesised policies satisfying probabilistic computation tree logic (PCTL) formulae. 
We consider several common benchmarks/problems and compare our work to recent developments for verifying upMDPs.
\end{abstract}
\begin{keywords}%
    Markov decision processes; Robust optimization; Verification; Scenario approach.
\end{keywords}

\section{Introduction}

Verifying the safety of complex dynamical systems is an important challenge \citep{knightSafetyCriticalSystems2002}, with applications including unmanned aerial vehicles (UAVs) \citep{yuDistributedMotionCoordination2021} and robotics \citep{DBLP:journals/arcras/BrunkeGHYZPS22}. 
Ensuring a safety property may require verifying satisfaction of complex and rich formal specifications in the process of uncertainty arising from, for example, inaccurate modelling or process noise.
A vital aspect of verification lies in finding abstractions that encompass this uncertainty, whilst accurately modelling system dynamics to aid optimal control of such systems.

In this paper we do not consider how such an abstraction may be generated and focus solely on learning a good controller/policy.
There are a number of techniques for abstracting dynamical systems, such as the widely celebrated counter-example guided abstraction/refinement approach \citep{clarkeVerificationHybridSystems2003}.
Alternatively, techniques make use of the so-called scenario approach in order to develop their models \citep{badingsRobustControlDynamical2023b, rickardFormalControllerSynthesis2023}.

One useful abstract model is that of parametric Markov decision processes (pMDPs) \citep{dawsSymbolicParametricModel2004, hahnProbabilisticReachabilityParametric2011, jungesParameterSynthesisMarkov2019a}, and particularly their probabilistic counterpart uncertain pMDPs (upMDPs) \citep{badingsScenariobasedVerificationUncertain2022, scheftelowitschMultiObjectiveApproachesMarkov2017}.
Parametric MDPs extend MDPs by parameterising their transition function; any choice of parameters induces a standard MDP.
Hence, a pMDP represents a family of MDPs, differing only in their transition function.
By drawing these parameters from a (possibly unknown) distribution, we define an uncertain parametric MDP. 
The parameterisation of the model allows for the introduction of some structure, thus avoiding overly conservative models. 
Here, we make no assumptions on the structure of the parameterisation, and could equivalently have a direct distribution over MDPs within an uncertain set, hence arriving at a problem closely related to that of robust MDPs \citep{iyengarRobustDynamicProgramming2005, nilimRobustControlMarkov2005, wiesemannRobustMarkovDecision2013}

A useful tool for expressing specifications on (up)MDPs lies in temporal logic \citep{platzerLogicsDynamicalSystems2012}, a rich language for specifying behaviours of a system \citep{beltaFormalMethodsDiscreteTime2017}. 
One particular language of interest is that of Probabilistic Computation Tree Logic (PCTL \citep{hanssonLogicReasoningTime1994}), an extension of Computation Tree Logic which allows for probabilistic quantification of properties. 
This language can be used to describe probabilistic specifications on system behaviour, with these probabilities allowing for uncertainty.
Often, it is of interest to learn a policy which ensures satisfaction of the specification \citep{hahnSynthesisPCTLParametric2011}, even under different realisations of the uncertainty. 

One common approach for verifying pMDPs is to solve the so-called \emph{parameter synthesis} problem, finding parameter sets that satisfy the specification. 
Typically, only a single set of parameters is of interest \citep{cubuktepeConvexOptimizationParameter2022, dawsSymbolicParametricModel2004a, meedeniyaEvaluatingProbabilisticModels2014}. 
Instead, we wish to synthesise policies that are probabilistically robust to uncertainty over the entire parameter space.

In this paper, we investigate the following problem.
Given a upMDP, with a possibly unknown distribution over parameters, we aim at synthesising a policy, accompanied by a probabilistic certificate, such that, for an MDP defined with a randomly drawn parameter set, the probability that the policy satisfies a given specification on that MDP is guaranteed by the computed certificate.
To solve this problem, we capitalize on advancements from the so-called \emph{scenario approach} literature \citep{calafioreScenarioApproachRobust2006,campiScenarioApproachSystems2009,DBLP:journals/mp/CampiG18,garattiRiskComplexityScenario2022a}. 
We sample a finite number of parameter sets, and compute a policy that is robust to any sample within that set. 
Then, by leveraging results from \cite{garattiRiskComplexityScenario2022a} we provide probably approximately correct (PAC) guarantees that the policy will be robust to a newly sampled parameter set. 
Importantly, these techniques allow us to provide guarantees based only on finite samples, without knowledge of the underlying distribution, and with no assumptions on the shape of this distribution or the geometry of its support set.


Since we are optimising a policy for multiple MDPs, our work is similar to developments on multiple environment MDPs (MEMDPs \citep{raskinMultipleEnvironmentMarkovDecision2014a, vegtRobustAlmostSureReachability2023}). 
These techniques consider finding policies which are optimal for a finite set of MDPs. 
In contrast to our approach, these methods assume to have complete knowledge of possible environments, whereas we wish to be robust to new, unseen, environments.
Previous work in this area typically required overly conservative assumptions on the parameterisation, restricting to models where the uncertainty is independent across different states \citep{puggelliPolynomialTimeVerificationPCTL2013a}, or across different state-action pairs \citep{kozineIntervalValuedFiniteMarkov2002}.
In the robust MDP literature, such assumptions are referred to as $s$-rectangular or $(s,a)$-rectangular ambiguity sets \citep{wiesemannRobustMarkovDecision2013}.
Some recent approaches \citep{badingsScenariobasedVerificationUncertain2022} have relaxed this assumption, but required full knowledge of the parameter sets at runtime in order to apply a policy.

Our work makes no assumption on the parameterisation of the upMDP, and learns a single policy which may be applied at runtime with no explicit parameter knowledge, whilst still offering PAC-type guarantees.
Our main contributions can be summarised as follows
\begin{enumerate}
    \item We formally discuss different policy classes for an MDP, and investigate their associated probabilistic guarantees. This policy classification is interesting per se as it clarifies subtle differences among them.
    \item We propose a new gradient-based algorithm to learn an optimal \emph{robust} policy for a given specification in an upMDP; we demonstrate the benefits of this scheme over more traditional solution paradigms via extensive numerical investigation and set the basis for a theoretical convergence analysis in future work.
    \item We use recent advancements in scenario approach theory to accompany our learned policy with non-trivial guarantees on the probability that newly sampled MDPs meet certain specifications.
\end{enumerate}

\section{Background}
\label{sec:background}

\subsection{Markov Decision Processes}

A Markov Decision Process (MDP) is a tuple $\MDP=(\states, \acts, \probs, \rho, \gamma)$. 
Where $\states = \{s_0, \dots, s_N \}$ is a finite set of states, with initial distribution $\rho:\states\rightarrow(0,1)$, $\acts = \{a_0, \dots, a_M \}$ a finite set of actions, $\probs \colon \states \times \acts \rightarrow \Dist(\states)$ is a probabilistic transition function, with $\Dist(\states)$ the set of all probability distributions over $\states$ \citep{baierPrinciplesModelChecking2008a}, and $\gamma\in(0,1)$ is some discount factor. 

We call a tuple $(s,a,s')$ with probability $\probs(s,a)(s') > 0$ a \emph{transition}.
By absorbing state, we refer to a state $s \in \states$ in which all transitions return to that state with probability 1 so that $\probs(s,a)(s) = 1$, for all $a \in \acts$, these typically being some goal or critical states.
An infinite trajectory of an MDP is a sampled sequence of states and actions $\traj = s_Ia_0s_1a_1\dots$, where actions are chosen according to some \emph{policy}, which we define in the sequel.

\subsubsection{Policy Classes}
\label{sec:background:MDP:pols}
We consider three distinct classes of policy to determine actions in this MDP.
Namely, deterministic (also called pure), mixed (also called randomized), and behavioural policies. 
These policies are denoted by $\pi^D \in \Pi^D, \pi^M \in \Pi^M, \pi^B \in \Pi^B$, referring to a deterministic, mixed and behavioural policy respectively. Specifically,
\begin{equation}
    \pi^D \colon \mathcal{S} \rightarrow \mathcal{A},\hspace{2cm}
    \pi^M = \text{Dist}(\Pi^D), \hspace{2cm} 
    \pi^B \colon\mathcal{S} \rightarrow \text{Dist}(\mathcal{A}).
\end{equation}
In other words, a deterministic policy shows which action to pick at each state; a mixed policy provides directly a distribution over deterministic policies, so that we sample one policy at the start of a trajectory and follow it throughout; a behavioural policy gives us a distribution over actions at each state, so that we sample one action at each state of a trajectory. 
For MDPs, it can be shown that deterministic policies suffice for optimality, but this doesn't hold for more complex models.

We denote by $\mixedProb$ the probability of choosing $\pi^D$ under the distribution defined for the mixed policy, and by $\behaviouralProb$ the probability of choosing action $a$ in a state $s$, under the policy defined by $\pi^B$.
Mixed and behavioural policies offer two semantically different options to resolve the non-determinism in a probabilistic manner. 
For MDPs, there \emph{is} a link between behavioural and mixed policies through Kuhn's theorem \citep{arrowContributionsTheoryGames1953}.
Further, for an MDP, it is always possible to find memoryless behavioural policies that are realization-equivalent to mixed policies (see 
\ifappendix
    \cref{app:proofs} for a proof).
\else 
    \cite[Appendix A]{TechRep} for a proof).
\fi
In Uncertain Parametric MDPs (introduced in the sequel) the link between behavioural and mixed policies may not be satisfied.

\subsubsection{Uncertain Parametric MDP}

An Uncertain Parametric MDP (upMDP) is a tuple $\upMDP = (\states, \acts, \params, \paramProbs, \probDist, \rho, \gamma)$.
Similar to an MDP, but the probabilistic transition function is now $\paramProbs \colon \states \times \acts \times \params \rightharpoonup \Dist(\states)$, so that we denote by $\paramProbs_v(s,a)(s')$ the probability of transitioning to state $s'$, given action $a$ in $s$, with parameters $v \in \params$ (i.e. the transition function is \emph{parameterised} by $v \in \params$), and with $\probDist$ defining a probability distribution over the parameter space.
Thus, when a set of parameters $v$ is sampled, we extract a concrete MDP ${\MDP}[v]$.
We do not impose any structure for this parameterisation, and only require that ${\MDP}[v]$ be a well-defined MDP. 
To uncover a trajectory in this upMDP we first sample a set of parameters $v$ from $\probDist$, and then follow a trajectory in the resulting MDP.
We denote by $\probDist^N$ the product measure associated with $N$ sampled parameter sets.

\subsection{Probabilistic Computation Tree Logic}

Probabilistic computation tree logic (PCTL) depends on the following syntax:
\begin{equation}
        \stateForm ::= \mathrm{true} \spacer p \spacer \neg \stateForm \spacer \stateForm \wedge \stateForm \spacer \PCTLp_{\sim \lambda}(\pathForm)\hspace{2cm}
        \pathForm ::= \stateForm\, \until\, \stateForm. 
\end{equation}
\noindent
Here, $\sim \in \{ <, \leq, \geq, > \}$ is a comparison operator and $\lambda \in [0,1]$ a probability threshold; PCTL formulae $\stateForm$ are state formulae, which can in particular depend on path formulae $\pathForm$.   
Informally, the syntax consists of state labels $p \in AP$ in a set of atomic propositions $AP$, propositional operators negation $\neg$ and conjunction $\wedge$, and temporal operator until $\until$.
The probabilistic operator $\PCTLp_{\sim \lambda}(\pathForm)$ requires that paths generated from the initial distribution satisfy a path formula $\pathForm$ with total probability exceeding (or below, depending on $\sim$) some given threshold $\lambda$. 
We consider only infinite horizon until, but note that extensions to include the finite time bounded-until operator can be achieved as discussed in \cite{DBLP:journals/jmlr/OsbandRRW19}. 

The satisfaction relation $\sampledMDP \sat_{\pol} \stateForm$ defines whether a PCTL formula $\stateForm$ holds true, when following policy $\pol$ in the concrete MDP. 
Formal definitions for semantics and model checking are provided in \cite{hanssonLogicReasoningTime1994,baierPrinciplesModelChecking2008a}.

\subsubsection{PCTL Satisfaction}

This satisfaction relation defines slightly semantics depending on the class of policy.
We now explore these semantics for the policy classes identified in \cref{sec:background:MDP:pols}, namely, deterministic, mixed and behavioural, respectively. 

    \begin{align}
            \MDP \sat_{\pi^D} \PCTLp_{\geq \lambda}(\pathForm) \impliedby& \sum_{\{\traj\colon \traj \sat \pathForm\}} \rho(s_0)\probs(s_1\given s_0, \pol^D(s_0))\probs(s_2\given s_1, \pol^D(s_1))\dots \geq \lambda
     \label{eq:PCTL_Det}\\ 
    &=\sum_{\{\traj\colon \traj \sat \pathForm\}} P_{\pol^D}(\traj)\geq \lambda, \nonumber\\
    \MDP \sat_{\pi^M} \PCTLp_{\geq \lambda}(\pathForm) \impliedby& \sum_{\pol^D\in\pols^D} \pol^M(\pol^D) \cdot \left( \sum_{\{\traj\colon \traj \sat \pathForm\}} P_{\pol^D}(\traj) \right)\geq \lambda,   \label{eq:PCTL_Mix}\\   
    \MDP \sat_{\pi^B} \PCTLp_{\geq \lambda}(\pathForm) \impliedby& \sum_{\{\traj\colon \traj \sat \pathForm\}} \rho(s_0)\probs(s_1\given s_0, a_0)\pol^B(a_0\given s_0)\probs(s_2\given s_1, a_1)\pol^B(a_1\given s_1)\dots  \geq \lambda\label{eq:PCTL_Behaviour} \\
    &=\sum_{\{\traj\colon \traj \sat \pathForm\}} P_{\pol^B}(\traj)\geq \lambda. \nonumber
    \end{align}

We use the $P_{\pol}(\traj)$ to refer to the probability of uncovering a trajectory when playing a given policy.
Note that both \cref{eq:PCTL_Mix,eq:PCTL_Behaviour} contain terms relating to the probability distributions introduced in the policy classes.
Instead, \cref{eq:PCTL_Det} only considers uncertainty in the model itself.

\subsection{Robust Policies}

We are interested in synthesising \emph{robust policies} for upMDPs.
That is, policies that maximise the probability of satisfying a PCTL formula $\pathForm$, within some given risk tolerance.
Thus, we are interested in solving the chance-constrained optimisation program
\begin{equation}
\label{eq:chance_prob}
            \max_{\pol \in \pols,~ \lambda \in [0,1]}  \lambda \hspace{0.7cm}
    \text{subject to }\hspace{0.3cm} \probDist\left\{ v \in \params \vert \sampledMDP \sat_{\pol} \PCTLp_{\geq \lambda}(\pathForm) \right\} \geq 1-\epsilon.
\end{equation}
for some a priori chosen risk level $\epsilon \in (0,1)$.
There is an inherent tradeoff here between risk and satisfaction probability: namely, for a small risk we might obtain a small satisfaction probability, and vice versa.
For brevity, we discuss only maximising $\lambda$, but note that the minimisation for $\sim \in \{<,\leq\}$ can be obtained trivially.
Note that here we optimise over policies $\pol$ in a generic set $\pols$; the exact policy class from the ones of \cref{sec:background:MDP:pols} will be specified in the sequel.
\section{Robust Policy Synthesis}
\label{sec:robust}

\begin{problem}
    \label{prob:formal}
    Given an upMDP $\upMDP$, PCTL formula $\pathForm$ and risk level $\epsilon$, find a robust policy $\pol$ and maximum satisfaction probability $\lambda$, such that, with probability $1-\epsilon$ playing policy $\pol$ on a newly sampled MDP will satisfy the PCTL formula with satisfaction probability at least $\lambda$.
\end{problem}

Due to the chance constraint on the parameter distribution $\probDist$, this problem is a semi-infinite optimization program (having a finite number of optimisation variables, but infinite constraints), and is generally intractable. 
Further, we may not have access to an analytical form for $\probDist$.
Thus, we turn to a sample based analogue to this problem.
In \cref{sec:PAC}, we investigate how solutions to this problem can provide guarantees to the original chance-constrained problem.


Consider an upMDP $\upMDP$ and $N$ i.i.d. sampled parameter sets (or \emph{scenarios}) $\allSamples=\{u_1,\dots,u_N \}$ sampled from $\probDist$. 
In this section, we investigate how to learn a robust policy which maximises the probability of satisfying a PCTL formula $\pathForm$, under the worst case realisation of the parameters, i.e. we investigate the following \emph{scenario program} associated to \cref{eq:chance_prob}
\begin{equation}
\label{eq:scen_prob}
            \max_{\pi \in \Pi,~ \lambda \geq 0} \lambda \hspace{1cm} 
    \text{subject to } \sol{\pi}{u} \geq \lambda, \; \forall u \in \allSamples.
\end{equation}
Where we use the notation $\sol{\pol}{u}$ to refer to the maximum probability of satisfying a formula $\pathForm$, in the concrete MDP $\sampledMDP$, under policy $\pol$, 
\begin{equation}
\label{eq:sol}
    \sol{\pol}{v} = \argmax_{\lambda \in [0,1]} \sampledMDP \sat_{\pol} \PCTLp_{\geq \lambda}(\pathForm).
\end{equation}

This scenario program problem coincides closely with the concept of Multiple Environment MDPs (MEMDPs) \citep{raskinMultipleEnvironmentMarkovDecision2014a}, but differs in some key ways.
When finding an optimal policy for MEMDPs, one wishes to find the policy that is optimal for the given environments.
Instead, our goal to find a policy that will be robust to an unseen environment. 
Furthermore, the concrete chance constrained problem we study is drastically different to MEMDPs since it involves continuous parameters rather than a finite set of environments.

It is shown in \cite{raskinMultipleEnvironmentMarkovDecision2014a} that infinite memory policies are required for optimality in MEMDPs.
In our setting, infinite memory policies will indeed lead to optimal satisfaction probabilities in the sample set, but suffer when generalising to new samples. 
This can be seen as a problem of overfitting, and may also lead to a deterioration in our guarantees. 

In the sequel, we provide different solution methodologies for \cref{eq:scen_prob}, where each methodology is based on a different class of policies.
\subsection{Solution by Interval MDPs (under deterministic policies)}
\label{sec:robust:iMDP}

A straightforward approach to solve this problem is to ignore the structure induced by the parameterisation, and to build instead an interval MDP, where each transition is only known up to an interval. 
Each interval can be constructed by looking at each transition in turn and finding the minimum and maximum probabilities from the sampled MDPs.
Such techniques are examined further (generally when no parameterisation is available) in \cite{kozineIntervalValuedFiniteMarkov2002} and are also closely aligned with the concept of $(s,a)$-rectangularity \citep{wiesemannRobustMarkovDecision2013}. The resulting policy is in general very conservative, since it considers the worst case probability for every single transition.
In reality, it is likely to be the case that the worst case transitions are unlikely to co-occur (for example, in a UAV motion planning problem like the one considered in \cref{tab:exp_res} the wind may push us left or right into an obstacle, but is unlikely to do both).

\subsection{MaxMin Game (under mixed policies)}
\label{sec:robust:Mixed}

The problem in \cref{eq:scen_prob} can be rewritten as a MaxMin problem $\max_{\pol \in \pols} \min_{u \in \allSamples} \sol{\pol}{u},$
which can be seen as a two player zero-sum game.
In which we have a policy player, and the samples act as an adversary. 
The policy player's actions are the set of all deterministic policies $\Pi^D$ for the upMDP, providing a finite, but potentially very large $\mathcal{O}(|\acts|^{|\states|} )$, action set.
The sample adversary's actions are the set of samples. 
Given sample $u$ and policy $\pol^D$, the reward to the policy player is $r_p(\pol^D, u) = \sol{\pol^D}{u}$.
We may solve this as a Stackelberg game \citep{stackelbergTheoryMarketEconomy1952,yousefimaneshStrategicRationingStackelberg2023}, with the adversary as the leader, to obtain an optimal deterministic policy.
Details on this method, and associated results, are available in 
\ifappendix
    \cref{app:det}.
\else
    \cite[Appendix E]{TechRep}.
\fi

Alternatively, a mixed strategy set $(s_p, s_\sigma)$, contains finite probability distributions over possible actions of each player, respectively, and returns the reward $r_{p}(s_{p}, s_\sigma) = \sum_{\pol^D \in \pols^D} \sum_{u \in \allSamples}s_{p}(\pi^D)\cdot s_\sigma(u)\cdot\sol{\pol_D}{u}$, or in matrix form $s_p R s_\sigma$, where matrix $R\in\mathbb{R}^{|\pols_D|\times N}$ has elements $R_{\pol_D^i, u_j}=\sol{\pol_D^i}{u_j}$.
Since the game consists of a finite number of players, each with a finite set of pure strategies, then by allowing players to play mixed strategies, it can be shown that a Nash equilibrium of the game will exist, and further, that all Nash equilibria have the same value.
Note that the mixed strategy defines a mixed policy $s_p = \pi^M$.

\begin{theorem}[Nash Equilibrium~\cite{nashNoncooperativeGames1989,DBLP:journals/tac/FrihaufKB12}]
\label{thm:NE}
For a two player zero-sum game, there exists at least one mixed strategy profile $s^\star = (s^\star_p,s^\star_\sigma)$, such that 
\begin{equation}
        r_p(s^\star_p, s^\star_{\sigma}) \geq r_p(s_p, s^\star_\sigma), \forall s_p \in S_p, \hspace{1cm}
        r_p(s^\star_p, s^\star_{\sigma}) \leq r_p(s^\star_p, s_\sigma), \forall s_\sigma \in S_\sigma.
\end{equation}
If both inequalities are strict, then there is a single unique Nash equilibrium, called a strict Nash equilibrium.
Otherwise, there is a set of Nash equilibria, all having equal value.
\end{theorem}

Finding Nash equilibrium strategies in this game is relatively straightforward, and there are a number of existing methods for solving the game (for example, the Porter-Nudelman-Shoham (PNS) algorithm \citep[Algorithm 1]{porterSimpleSearchMethods2008} or fictitious play methods \citep{heinrichFictitiousSelfPlayExtensiveForm2015}).
Thus, we can simply build the reward matrix $R$, by considering each determinsitic policy and sample in turn, and pass this to one of these algorithms.
Unfortunately, this method is computationally very expensive, with the size of the reward matrix being $\mathbb{R}^{N\times|\states|^{|\acts|}}$, and algorithms to solve such games being exponential in the size of this matrix. 

\subsection{Subgradient Ascent (under behavioural policies)}
\label{sec:robust:sub}
Under the choice of behavioural policies, we propose an alternative method that avoids computing a full payoff matrix for every deterministic policy.
Taking inspiration from \cite{DBLP:journals/corr/abs-1906-01786}, we rewrite the solution function as $\sol{\pol^B}{u} = \sum_{s \in \states} \eta^u_{\pi^B}(s) \sum_{a \in \acts} Q^u_{\pol^B}(s,a)\pol^B(s)(a),$
where $Q^u_{\pol^B}(s,a)$ is a Q-function, defined to model the PCTL requirement 
\ifappendix
    (see \cref{app:PCTL_Q}) 
\else
    \citep[Appendix B]{TechRep} 
\fi
and $\eta^u_{\pi^B}(s)$ is the discounted state-occupancy measure $\eta^u_{\pi^B}(s) = (1-\gamma)\sum_{k=0}^\infty \gamma_k P^{\pol^B}_k(s)$, with $P^{\pol^B}_k(s)$ the probability of uncovering state $s$ at time $k$.
Intuitively, $\eta^u_{\pi^B}(s)$ defines the (discounted) fraction of time the system spends in a given state. 
We fix $\eta^u_{\pol^B}$ and $Q^u_{\pol^B}$ (as in policy iteration \citep{fosterDynamicProgrammingMarkov1962b}), and find the gradient (given analytically in \cref{alg:sub:grad}).

This provides us with a (sub)gradient for a single solution function, however, we are primarily interested in the pointwise minimum amongst a set of solution functions. 
Hence, we turn to subdifferentials (or \emph{subgradients}) \citep{kiwielConvergenceEfficiencySubgradient2001, ausselMeanValueProperty1995}. 
One simple way of finding a valid subdifferential is to find the gradient of one of the current minimum functions.

\begin{algorithm}
\caption{Projected Subgradient Ascent}
\label{algo:sub}
\KwData{upMDP $\MDP$, samples $\allSamples$, formula $\pathForm$, step-size sequence $\{\alpha_0,\alpha_1,\dots\}$}
\KwResult{Optimal Policy $\pol$}
    $k\gets0$ \quad
    $\pi^B_0 \gets \text{uniform random}$\\
    \While{not converged}{
        $u^\star \gets \text{random choice}(\argmin_{u \in \allSamples} \sol{\pol^B_k}{u})$ \hfill \tcp{Select worst case sample}
        $\nabla_{k+1}(s,a) = \eta^{u^\star}_{\pi^B}(s) Q^{u^\star}_{\pol^B}(s,a)$ \label{alg:sub:grad} \hfill \tcp{Find gradient for worst case} 
        $\pi^B_{k+1}(s)(a) \gets \text{proj}_{\sum \pi^B(\cdot \given s)=1}[\pi^B_k+\alpha_k \nabla_{k+1}(s,a)]$ \hfill \tcp{Gradient step}
        $f^{k+1}_\ast = \max\{f^k_\ast, f(\pi^B_{k+1}) \}$ \hfill \tcp{Store record objective}
        $k \leftarrow k+1$\\
    }
\end{algorithm}%
We initialise with a uniform random policy, select a minimising (worst-case) sample from the set of minimisers (which may not be a singleton), then find the gradient for this sample, take a step in this direction, and project onto the constraint set.
We use a diminishing, non-summable step size $\alpha_k$, typically employed in subgradient methods.
We give numerical evidence that this algorithm exhibits a convergent behaviour in \cref{sec:results}, and further analysis in 
\ifappendix
    \cref{app:add_res}.
\else
    \cite[Appendix D]{TechRep}.
\fi
\section{Guarantees}
\label{sec:PAC}

Once we have synthesised a policy using one of the methods above, we can accompany each of the policies synthesized according to the aforementioned methodologies with PAC-type guarantees on the satisfaction of the PCTL property under consideration.
To achieve this, we use the following recent result in the scenario approach theory.


Given the solution to a scenario program, we are interested in quantifying the risk associated with the solution (i.e. the probability that our solution violates a new sampled constraint).
The scenario approach provides us with the techniques to achieve this by considering the number of support constraints.
In an optimisation problem, if the removal of a constraint leads to a changed solution, then this constraint is said to be a support constraint.
Further, if a solution returned when considering only the support constraints is the same as the solution obtained when employing all samples, then the problem is a non-degenerate problem.

\begin{theorem}[PAC Guarantees \cite{garattiRiskComplexityScenario2022a}]
    Consider the optimisation problem in \cref{eq:scen_prob}, with $N$ sampled parameter sets.
    Given a confidence parameter $\beta \in (0,1)$, for any $k = 0, 1, \dots, N$, consider the polynomial equation in the variable $t$
    \begin{equation}
    \label{eq:risk}
    \xi_k(t) = \begin{cases}
                1-\frac{\beta}{6N}\sum_{i=N+1}^{4N}\binom{i}{k}t^{i-k}, & \text{if } k = N,\\
                \binom{N}{k}t^{N-K}-\frac{\beta}{2N}\sum_{i=k}^{N-1}\binom{i}{k}t^{i-k}-\frac{\beta}{6N}\sum_{i=N+1}^{4N}\binom{i}{k}t^{i-k}, & \text{otherwise.} \\
    \end{cases}
    \end{equation}
    Solving $\xi_k(t) = 0$ for $t \in [0,+\infty)$, for $k = 0, 1, \dots, N-1$, we find exactly two solutions, which we denote with $\underline{t}(k), \overline{t}(k)$ with $\underline{t}(k) \leq \overline{t}(k)$.
    For $k=N$, we find a single solution, which we denote by $\overline{t}(N)$, and define $\underline{t}(N) = 0$. 
    Let $\underline{\epsilon}(k) \coloneqq \max\{0, 1-\overline{t}(k)\}$ and $\overline{\epsilon}(k) \coloneqq 1-\underline{t}(k)$.
    
    Assume the problem is non-degenerate\footnote{Non-degeneracy implies that solving the problem using only the samples that are of support results in the same solution (policy in our case) had all the samples been employed. We say that a sample (which gives rise to a constraint in \cref{eq:scen_prob}) is of support, if removing only that sample results in a different solution. Repeating this procedure, removing samples one-by-one allows identifying the support samples of a given problem.}, and has a unique solution. Let $\pol_N^\ast$ be the optimal policy, $\lambda_N^\ast$ the optimal satisfaction probability,
    and $\Tilde{s}^\ast_N$ is the number of support samples. 
    Then, for any $\mathbb{P}$ it holds that
    \begin{equation}
        \mathbb{P}^N\{\underline{\epsilon}(\Tilde{s}^\ast_N) \leq \mathbb{P}\left\{v \in \params \colon \sampledMDP \notsat_{\pi^M} \PCTLp_{\geq \lambda^\star}(\pathForm) \right\}\leq \overline{\epsilon}(\Tilde{s}^\ast_N) \} \geq 1-\beta,
    \end{equation}
    where $\mathbb{P}\left\{v \in \params \colon \sampledMDP \notsat_{\pi^M} \PCTLp_{\geq \lambda^\star}(\pathForm) \right\}$ is the risk (the probability that our solution violates the specification for another sample $v$). 
\end{theorem}

Note that in $\pol_N^\ast$ we do not specify the problem class; this is considered in the sequel when we deploy this theorem for each solution methodology from the previous section. Moreover, to determine the 
number of support samples $\Tilde{s}^\ast_N$, we exploit our problem's structure and, based on the solution methodology adopted, we discuss how an upper-bound on their number can be obtained.

To this end, we provide guarantees for the policy generated by each of the methods of Section \ref{sec:robust}.
Consider first the case of the mixed policy determined using a Nash equilibrium solver as per the developments of Section \ref{sec:robust:Mixed}.
In this case, finding the number of support samples consists of finding samples which are included in the mixed strategy of the sampled adversary.
\begin{corollary}[PAC Guarantees for Mixed Policies]
Consider the MaxMin game of Section \ref{sec:robust:Mixed}, and let $\pol_N^\ast$ be the optimal mixed policy $\pi^M$ returned by a Nash equilibrium solver.
The number of support constraints may be found as $\supps = |\{u \in \allSamples \colon s_\sigma(u) > 0\}|$, while the satisfaction relation $\sampledMDP \notsat_{\pi^M} \PCTLp_{\geq \lambda^\star}(\pathForm)$ is as defined in \cref{eq:PCTL_Mix}.
\end{corollary}

Consider now the case of a behavioural policy obtained using the subgradient methodology (\cref{sec:robust:sub}). 
We determine the number of support constraints by finding the active constraints (as the problem is assumed to be non-degenerate, the active constraints are also the support constraints). 
\begin{corollary}[PAC Guarantees for Behavioural Policies]
Consider the subgradient algorithm of Section \ref{sec:robust:Mixed}, and let $\pol_N^\ast$ be the optimal behavioural policy $\pi^B$ returned by Algorithm \ref{algo:sub}.
    The number of support samples is given by the active constraints, which are in turn the ones for which the obtained satisfaction probability $\lambda^\star$ is tight, i.e., $\supps = |\{u \in \allSamples \colon \sol{\pol^B}{u} = \lambda^\star\}|,$
    while the satisfaction relation $\sampledMDP \notsat_{\pi^B} \PCTLp_{\geq \lambda^\star}(\pathForm)$ is as defined in \cref{eq:PCTL_Behaviour}.
\end{corollary}

Finally, consider the construction of an interval MDP under the class of deterministic policies.
In this case, our problem is degenerate, since it may be necessary to remove multiple constraints for another policy to become optimal, thus prohibiting the use of Theorem 2. 
Therefore, we leverage techniques from \cite{campiGeneralScenarioTheory2018} that do not require imposing a non-degeneracy assumption. 
 
\begin{corollary}[PAC Guarantees for iMDP Policies]
   Fix $\beta \in (0,1)$, and as in \cite[Theorem 1]{campiGeneralScenarioTheory2018}, define $\mu(N) \coloneqq 1$, and for $\supps<N$ let
   \begin{equation}
       \label{eq:det_risk}
       \mu(\supps) \coloneqq 1 - \sqrt[N-\supps]{\frac{\beta}{N \binom{N}{\supps}}}.
   \end{equation}
    Support samples are those which define the intervals: $\overline{\supps} = |\{u \in \allSamples \colon \exists (s,a,s'),  \probs_u(s,a)(s') \leq \probs_v(s,a)(s')
            \vee \probs_u(s,a)(s') \geq \probs_v(s,a)(s')
            ,  \forall v \in \allSamples\}|,$
            i.e., those with at least one transition probability at an extremum of its interval.
    Then (with satisfaction relation defined in \cref{eq:PCTL_Det}) we have 
\begin{align}
        \mathbb{P}^N\left\{\mathbb{P}\left\{v \in \params \colon \sampledMDP \notsat_{\pi^D} \PCTLp_{\geq \lambda^\star}(\pathForm) \right\} \leq \mu(\overline{\supps})\right\} \geq 1-\beta.
\end{align}
\end{corollary}
\section{Numerical Experiments}
\label{sec:results}

We implemented our techniques in Python and made use of the probabilistic model checker Storm \citep{henselProbabilisticModelChecker2022} to verify PCTL formulae on MDPS, and PRISM \citep{kwiatkowskaPRISMVerificationProbabilistic2011} for iMDPs. 
Experiments were run on a server with 80 2.5 GHz CPUs and 125 GB of RAM. 
The codebase is available at \href{https://github.com/lukearcus/robust\_upMDP}{https://github.com/lukearcus/robust\_upMDP}  

We evaluate our techniques on a number of benchmark models available in the literature \citep{DBLP:conf/atva/QuatmannD0JK16}, as well as a few simpler examples, for a complete description see
\ifappendix
    \cref{app:exp_det}.
\else
    \cite[Appendix C]{TechRep}.
\fi
We compare our subgradient algorithm (\cref{sec:robust:sub}) to existing techniques from \cite{badingsScenariobasedVerificationUncertain2022} (column 1) and an implementation of the synthesis based on iMDP,  as in \cite{kozineIntervalValuedFiniteMarkov2002} (column 2).
Unless otherwise stated, we use a numerical tolerance of $10^{-4}$, a confidence parameter of $\beta=10^{-5}$, take $N=200$ parameter samples, and allow 1 hour of computation time.
We provide numerical values for the optimal satisfaction probability $\lambda^\star$, the theoretical risk upper bound $\overline{\epsilon}$, the runtimes, the empirical values of the risk $\Tilde{\epsilon}$, and the empirical values of the risk without access to the true parameter set (if parameter access is needed to synthesise a policy, we draw a sample $v_1 \sim \probDist$ for synthesis, but test on sample, $v_2\sim\probDist$). 
We highlight in grey cases where we discuss next how the methodologies under comparison are compared/outperformed by our proposed subgradient algorithm either in terms of the quality of their probabilistic guarantees and/or the computational requirements, and use red text for empirical values which violate a bound.

Our subgradient algorithm is less conservative than a naive iMDP solution, 
offering non-trivial theoretical risk bounds (for the iMDP approach these bounds are often equal to one), and superior satisfaction probabilities (the iMDP approach considers the worst case probability for every transition).
By superior, we mean either higher or lower depending on the optimisation direction (maximisation or minimisation, respectively, denoted by (max) and (min) in \cref{tab:exp_res}).
The approach in \cite{badingsScenariobasedVerificationUncertain2022} generally outperforms our methods in speed, risk bounds and satisfaction probability. 
The difference on the theoretical risk bounds lies in the fact that our approach involves solving a problem with a non-trivial number of support constraints, while in \cite{badingsScenariobasedVerificationUncertain2022} they have only one support constraint. 
As such, it offers tighter guarantees, however, these refer to an existential statement. 
To see this, note that \cite{badingsScenariobasedVerificationUncertain2022} compute a different policy per sample, and provide guarantees only on the existence of a policy for a new sample. 
On the contrary, we construct a policy that is robust with respect to all samples, and at the same time is accompanied by guarantees on its feasibility properties for unseen samples. 
Moreover, \cite{badingsScenariobasedVerificationUncertain2022} 
require access to true parameters at runtime, and a non-trivial amount of online computation to solve the MDP.
Without this, their risk bounds may be violated (as seen in red). 

Our mixed policy solution (\cref{sec:robust:Mixed}), timed out on all but the toy model, for which the results are: Sat. Prob: .325, Risk U.B.: .147, runtime: 0.75s Emp. risk: .003, Emp. risk (no pars): .101.
\begin{table}[t]
    \centering
    \addtolength{\tabcolsep}{-0.3em}
    \resizebox{\textwidth}{!}{\begin{tabular}{p{1.25cm} x{1.5cm} c|p{0.7cm} p{0.75cm} p{0.7cm} p{0.7cm} p{0.7cm}|p{0.7cm} p{0.75cm} p{0.7cm} p{0.7cm} p{0.7cm}|p{0.7cm} p{0.75cm} p{0.82cm} p{0.7cm} p{0.7cm}}
        \hline
          & & & \multicolumn{5}{|c|}{\cite{badingsScenariobasedVerificationUncertain2022}}&\multicolumn{5}{|c|}{iMDP Solver} & \multicolumn{5}{|c|}{Subgradient   (\cref{algo:sub})}\\ \cline{4-18} 
         Model & Instance & $|\states|\cdot|\acts|$ & Sat. Prob. $\lambda^\star$ & Risk U.B. $\overline{\epsilon}$ & Time (s) & Emp. Risk $\Tilde{\epsilon}$ & Emp. (no pars) & Sat. Prob. $\lambda^\star$ & Risk U.B. $\overline{\mu}$ & Time (s) & Emp. Risk $\Tilde{\mu}$ & Emp. (no pars)& Sat. Prob. $\lambda^\star$ & Risk U.B. $\overline{\epsilon}$ & Time (s) & Emp. Risk $\Tilde{\epsilon}$ & Emp. (no pars)  \\ \hline \hline 
                    \multirow{3}{1cm}{\textbf{consensus} (min)}  & (2,2) & 306 & .967 & .056 & 1 & .004 & .008 & .967 & \cellcolor[gray]{0.8} 1 & 3 & .007 & .008  &.967 & .187 & 4 & .009 & .011\\
          &(2,32) & 5 586 & .996 & .056 & 136 & .000 & \cellcolor[gray]{0.8}.014 & .996 & \cellcolor[gray]{0.8}1 & 76 & .000 & .000 & .996 & .516 & 202 & .002 & .000 \\
          &(4,2) & 90 624 & - & - & TO & - & - & .936 & 1 & 1887 & .000 & .000 & - & - & TO & - & -  \\
         \hline
        \multirow{1}{2cm}{\textbf{brp} (max)}&(256,5) & 42 064 & \cellcolor[gray]{0.8}- & \cellcolor[gray]{0.8}- & \cellcolor[gray]{0.8}TO & \cellcolor[gray]{0.8}- & \cellcolor[gray]{0.8}- & .985 & 1 & 383 & .001 & .000 & .985 & 1 & 923 &.000 & .001   \\
         \hline 
        \multirow{3}{1.25cm}{\textbf{sav} (min)}&(6,2,2) & 1516 & .834 & .056 & 4 & .015 & \cellcolor[gray]{0.8}\textcolor{red}{.112} & \cellcolor[gray]{0.8}.923 & \cellcolor[gray]{0.8}1 & 9 & .002 & .004 & .884 & .180 & 218 & .013 & .007 \\
          &(6,2,2) & 1516 & .700 & .056 & 5 & .000 & .017 & \cellcolor[gray]{0.8}.721 & \cellcolor[gray]{0.8}1 & 9 & .017 & .010 & .720 & .187 & 76 & .011 & .017 \\
         &(10,3,3) & 7400 & .446 & .056 & 51 & .009 & \cellcolor[gray]{0.8}\textcolor{red}{.120} & .524 & \cellcolor[gray]{0.8} 1 & 50 &  .017 & .015 & .526 & .147 & 109 & .021 & .018  \\
         \hline
         \multirow{3}{1.25cm}{\textbf{UAV} (max)}&uniform & 45 852 & .301 & .056 & 978 & .003 & \cellcolor[gray]{0.8}\textcolor{red}{.749} & \cellcolor[gray]{0.8}.065 & \cellcolor[gray]{0.8}1 & 238 & .015 & .013 & .195 & .245 & 85719\footnotemark & .077 & .089 \\
         &x-neg & 45 852 & .198 & .056 & 804 & .002 & \cellcolor[gray]{0.8}\textcolor{red}{.802} & \cellcolor[gray]{0.8}.012 & \cellcolor[gray]{0.8}1 & 210 & .000 & .000& .102 & .440 & 58612 & .068 & .071\\
         &y-pos & 45 852 & .564 & .056 & 811 & .003 & \cellcolor[gray]{0.8}\textcolor{red}{.755} & \cellcolor[gray]{0.8}.050 & \cellcolor[gray]{0.8}1&196&.001&.001 & .444 & .163 & 92440 & .095 & .101 \\
         \hline
         \multicolumn{2}{p{3cm}}{\textbf{Toy Model} (max)} & 8 & .326 & .056 & .11 & .008 & \cellcolor[gray]{0.8}\textcolor{red}{.106} & \cellcolor[gray]{0.8}.323 & .155 & 1.47 & .008 & .094 & .325 & .366 & 8.95 & .003 & .098 \\
         \hline
    \end{tabular}}%
    \caption{Experimental Results; the results for the case of the MaxMin Solution are discussed in the text. TO indicates timed-out.}
    \label{tab:exp_res}
\end{table}
\footnotetext{For the UAV benchmark, we allowed up to 2 days of computation time. }

\section{Concluding Remarks and Future Directions}
\label{sec:conclusion}

We presented a novel method for learning a single robust policy for upMDPs, providing PAC guarantees on satisfaction of PCTL formulae. 
We have considered several policy classes, and provided guarantees for each. 
Our experiments demonstrate the efficacy of our methods on a number of benchmarks, and provide comparisons to previous methods.

One avenue for future work is discarding constraints \citep{campiSamplingandDiscardingApproachChanceConstrained2011b}.
Relatedly, improving the runtimes of our algorithms (perhaps at the cost of guarantees) is of interest.
Finally, we leave a full technical analysis of the convergence of our subgradient method to later work.




\acks{This work was supported by the EPSRC Centre for Doctoral Training in Autonomous Intelligent Machines and Systems EP/S024050/1.}


\bibliography{upMDP_paper}

\begin{thebibliography}{47}
\providecommand{\natexlab}[1]{#1}
\providecommand{\url}[1]{\texttt{#1}}
\expandafter\ifx\csname urlstyle\endcsname\relax
  \providecommand{\doi}[1]{doi: #1}\else
  \providecommand{\doi}{doi: \begingroup \urlstyle{rm}\Url}\fi

\bibitem[Arrow et~al.(1953)Arrow, Barankin, Blackwell, Bott, Dalkey, Dresher,
  Gale, Gillies, Glicksberg, Gross, Karlin, Kuhn, Mayberry, Milnor, Motzkin,
  Von~Neumann, Raiffa, Shapley, Shiffman, Stewart, Thompson, and
  Thrall]{arrowContributionsTheoryGames1953}
K.~J. Arrow, E.~W. Barankin, D.~Blackwell, R.~Bott, N.~Dalkey, M.~Dresher,
  D.~Gale, D.~B. Gillies, I.~Glicksberg, O.~Gross, S.~Karlin, H.~W. Kuhn, J.~P.
  Mayberry, J.~W. Milnor, T.~S. Motzkin, J.~Von~Neumann, H.~Raiffa, L.~S.
  Shapley, M.~Shiffman, F.~M. Stewart, G.~L. Thompson, and R.~M. Thrall.
\newblock \emph{Contributions to the {{Theory}} of {{Games}} ({{AM-28}}),
  {{Volume II}}}.
\newblock {Princeton University Press}, 1953.
\newblock ISBN 978-0-691-07935-6.

\bibitem[Aussel et~al.(1995)Aussel, Corvellec, and
  Lassonde]{ausselMeanValueProperty1995}
Didier Aussel, Jean-Noel Corvellec, and Marc Lassonde.
\newblock Mean {{Value Property}} and {{Subdifferential Criteria}} for {{Lower
  Semicontinuous Functions}}.
\newblock \emph{Transactions of the American Mathematical Society},
  347\penalty0 (10):\penalty0 4147--4161, 1995.
\newblock ISSN 0002-9947.

\bibitem[Badings et~al.(2022)Badings, Cubuktepe, Jansen, Junges, Katoen, and
  Topcu]{badingsScenariobasedVerificationUncertain2022}
Thom~S. Badings, Murat Cubuktepe, Nils Jansen, Sebastian Junges, Joost-Pieter
  Katoen, and Ufuk Topcu.
\newblock Scenario-based verification of uncertain parametric {{MDPs}}.
\newblock \emph{Int. J. Softw. Tools Technol. Transf.}, 24\penalty0
  (5):\penalty0 803--819, 2022.

\bibitem[Badings et~al.(2023)Badings, Romao, Abate, Parker, Poonawala,
  Stoelinga, and Jansen]{badingsRobustControlDynamical2023b}
Thom~S. Badings, Licio Romao, Alessandro Abate, David Parker, Hasan~A.
  Poonawala, Mari{\"e}lle Stoelinga, and Nils Jansen.
\newblock Robust {{Control}} for {{Dynamical Systems}} with {{Non-Gaussian
  Noise}} via {{Formal Abstractions}}.
\newblock \emph{J. Artif. Intell. Res.}, 76:\penalty0 341--391, 2023.

\bibitem[Baier and Katoen(2008)]{baierPrinciplesModelChecking2008a}
Christel Baier and Joost-Pieter Katoen.
\newblock \emph{Principles of Model Checking}.
\newblock {MIT Press}, 2008.
\newblock ISBN 978-0-262-02649-9.

\bibitem[Belta et~al.(2017)Belta, Yordanov, and
  Aydin~Gol]{beltaFormalMethodsDiscreteTime2017}
Calin Belta, Boyan Yordanov, and Ebru Aydin~Gol.
\newblock \emph{Formal {{Methods}} for {{Discrete-Time Dynamical Systems}}},
  volume~89 of \emph{Studies in {{Systems}}, {{Decision}} and {{Control}}}.
\newblock {Springer International Publishing}, {Cham}, 2017.
\newblock ISBN 978-3-319-50762-0 978-3-319-50763-7.
\newblock \doi{10.1007/978-3-319-50763-7}.

\bibitem[Bhandari and Russo(2019)]{DBLP:journals/corr/abs-1906-01786}
Jalaj Bhandari and Daniel Russo.
\newblock Global optimality guarantees for policy gradient methods.
\newblock \emph{CoRR}, abs/1906.01786, 2019.

\bibitem[Brunke et~al.(2022)Brunke, Greeff, Hall, Yuan, Zhou, Panerati, and
  Schoellig]{DBLP:journals/arcras/BrunkeGHYZPS22}
Lukas Brunke, Melissa Greeff, Adam~W. Hall, Zhaocong Yuan, Siqi Zhou, Jacopo
  Panerati, and Angela~P. Schoellig.
\newblock Safe learning in robotics: {{From}} learning-based control to safe
  reinforcement learning.
\newblock \emph{Annu. Rev. Control. Robotics Auton. Syst.}, 5:\penalty0
  411--444, 2022.

\bibitem[Calafiore and Campi(2006)]{calafioreScenarioApproachRobust2006}
Giuseppe~Carlo Calafiore and Marco~C. Campi.
\newblock The scenario approach to robust control design.
\newblock \emph{IEEE Trans. Autom. Control.}, 51\penalty0 (5):\penalty0
  742--753, 2006.

\bibitem[Campi and
  Garatti(2011)]{campiSamplingandDiscardingApproachChanceConstrained2011b}
Marco~C. Campi and Simone Garatti.
\newblock A {{Sampling-and-Discarding Approach}} to {{Chance-Constrained
  Optimization}}: {{Feasibility}} and {{Optimality}}.
\newblock \emph{J. Optim. Theory Appl.}, 148\penalty0 (2):\penalty0 257--280,
  2011.

\bibitem[Campi and Garatti(2018)]{DBLP:journals/mp/CampiG18}
Marco~C. Campi and Simone Garatti.
\newblock Wait-and-judge scenario optimization.
\newblock \emph{Math. Program.}, 167\penalty0 (1):\penalty0 155--189, 2018.

\bibitem[Campi et~al.(2009)Campi, Garatti, and
  Prandini]{campiScenarioApproachSystems2009}
Marco~C. Campi, Simone Garatti, and Maria Prandini.
\newblock The scenario approach for systems and control design.
\newblock \emph{Annu. Rev. Control.}, 33\penalty0 (2):\penalty0 149--157, 2009.
\newblock \doi{10.1016/j.arcontrol.2009.07.001}.

\bibitem[Campi et~al.(2018)Campi, Garatti, and
  Ramponi]{campiGeneralScenarioTheory2018}
Marco~Claudio Campi, Simone Garatti, and Federico~Alessandro Ramponi.
\newblock A {{General Scenario Theory}} for {{Nonconvex Optimization}} and
  {{Decision Making}}.
\newblock \emph{IEEE Trans. Autom. Control.}, 63\penalty0 (12):\penalty0
  4067--4078, 2018.

\bibitem[Clarke et~al.(2003)Clarke, Fehnker, Han, Krogh, Stursberg, and
  Theobald]{clarkeVerificationHybridSystems2003}
Edmund~M. Clarke, Ansgar Fehnker, Zhi Han, Bruce~H. Krogh, Olaf Stursberg, and
  Michael Theobald.
\newblock Verification of {{Hybrid Systems Based}} on {{Counterexample-Guided
  Abstraction Refinement}}.
\newblock In Hubert Garavel and John Hatcliff, editors, \emph{Tools and
  {{Algorithms}} for the {{Construction}} and {{Analysis}} of {{Systems}}, 9th
  {{International Conference}}, {{TACAS}} 2003, {{Held}} as {{Part}} of the
  {{Joint European Conferences}} on {{Theory}} and {{Practice}} of
  {{Software}}, {{ETAPS}} 2003, {{Warsaw}}, {{Poland}}, {{April}} 7-11, 2003,
  {{Proceedings}}}, volume 2619 of \emph{Lecture {{Notes}} in {{Computer
  Science}}}, pages 192--207. {Springer}, 2003.
\newblock \doi{10.1007/3-540-36577-X\_14}.

\bibitem[Cubuktepe et~al.(2022)Cubuktepe, Jansen, Junges, Katoen, and
  Topcu]{cubuktepeConvexOptimizationParameter2022}
Murat Cubuktepe, Nils Jansen, Sebastian Junges, Joost-Pieter Katoen, and Ufuk
  Topcu.
\newblock Convex {{Optimization}} for {{Parameter Synthesis}} in {{MDPs}}.
\newblock \emph{IEEE Trans. Autom. Control.}, 67\penalty0 (12):\penalty0
  6333--6348, 2022.

\bibitem[Daws(2004{\natexlab{a}})]{dawsSymbolicParametricModel2004}
Conrado Daws.
\newblock Symbolic and {{Parametric Model Checking}} of {{Discrete-Time Markov
  Chains}}.
\newblock In \emph{{{ICTAC}}}, volume 3407 of \emph{Lecture {{Notes}} in
  {{Computer Science}}}, pages 280--294. {Springer}, 2004{\natexlab{a}}.

\bibitem[Daws(2004{\natexlab{b}})]{dawsSymbolicParametricModel2004a}
Conrado Daws.
\newblock Symbolic and {{Parametric Model Checking}} of {{Discrete-Time Markov
  Chains}}.
\newblock In \emph{{{ICTAC}}}, volume 3407 of \emph{Lecture {{Notes}} in
  {{Computer Science}}}, pages 280--294. {Springer}, 2004{\natexlab{b}}.

\bibitem[Foster(1962)]{fosterDynamicProgrammingMarkov1962b}
F.~G. Foster.
\newblock Dynamic {{Programming}} and {{Markov Processes}}. {{By R}}. {{A}}.
  {{Howard}}. {{Pp}}. 136. 46s. 1960. ({{John Wiley}} and {{Sons}},
  {{N}}.{{Y}}.).
\newblock \emph{The Mathematical Gazette}, 46\penalty0 (358):\penalty0
  340--341, December 1962.
\newblock ISSN 0025-5572, 2056-6328.

\bibitem[Frihauf et~al.(2012)Frihauf, Krstic, and
  Basar]{DBLP:journals/tac/FrihaufKB12}
Paul Frihauf, Miroslav Krstic, and Tamer Basar.
\newblock Nash equilibrium seeking in noncooperative games.
\newblock \emph{{IEEE} Trans. Autom. Control.}, 57\penalty0 (5):\penalty0
  1192--1207, 2012.

\bibitem[Garatti and Campi(2022)]{garattiRiskComplexityScenario2022a}
Simone Garatti and Marco~C. Campi.
\newblock Risk and complexity in scenario optimization.
\newblock \emph{Math. Program.}, 191\penalty0 (1):\penalty0 243--279, 2022.
\newblock \doi{10.1007/s10107-019-01446-4}.

\bibitem[Hahn et~al.(2011{\natexlab{a}})Hahn, Han, and
  Zhang]{hahnSynthesisPCTLParametric2011}
Ernst~Moritz Hahn, Tingting Han, and Lijun Zhang.
\newblock Synthesis for {{PCTL}} in {{Parametric Markov Decision Processes}}.
\newblock In Mihaela~Gheorghiu Bobaru, Klaus Havelund, Gerard~J. Holzmann, and
  Rajeev Joshi, editors, \emph{{{NASA Formal Methods}} - {{Third International
  Symposium}}, {{NFM}} 2011, {{Pasadena}}, {{CA}}, {{USA}}, {{April}} 18-20,
  2011. {{Proceedings}}}, volume 6617 of \emph{Lecture {{Notes}} in {{Computer
  Science}}}, pages 146--161. {Springer}, 2011{\natexlab{a}}.
\newblock \doi{10.1007/978-3-642-20398-5\_12}.

\bibitem[Hahn et~al.(2011{\natexlab{b}})Hahn, Hermanns, and
  Zhang]{hahnProbabilisticReachabilityParametric2011}
Ernst~Moritz Hahn, Holger Hermanns, and Lijun Zhang.
\newblock Probabilistic reachability for parametric {{Markov}} models.
\newblock \emph{Int. J. Softw. Tools Technol. Transf.}, 13\penalty0
  (1):\penalty0 3--19, 2011{\natexlab{b}}.

\bibitem[Hansson and Jonsson(1994)]{hanssonLogicReasoningTime1994}
Hans Hansson and Bengt Jonsson.
\newblock A logic for reasoning about time and reliability.
\newblock \emph{Formal Aspects of Computing}, 6\penalty0 (5):\penalty0
  512--535, September 1994.
\newblock ISSN 1433-299X.
\newblock \doi{10.1007/BF01211866}.

\bibitem[Heinrich et~al.(2015)Heinrich, Lanctot, and
  Silver]{heinrichFictitiousSelfPlayExtensiveForm2015}
Johannes Heinrich, Marc Lanctot, and David Silver.
\newblock Fictitious {{Self-Play}} in {{Extensive-Form Games}}.
\newblock In \emph{{{ICML}}}, volume~37 of \emph{{{JMLR Workshop}} and
  {{Conference Proceedings}}}, pages 805--813. {JMLR.org}, 2015.

\bibitem[Hensel et~al.(2022)Hensel, Junges, Katoen, Quatmann, and
  Volk]{henselProbabilisticModelChecker2022}
Christian Hensel, Sebastian Junges, Joost-Pieter Katoen, Tim Quatmann, and
  Matthias Volk.
\newblock The probabilistic model checker {{Storm}}.
\newblock \emph{Int. J. Softw. Tools Technol. Transf.}, 24\penalty0
  (4):\penalty0 589--610, 2022.

\bibitem[Iyengar(2005)]{iyengarRobustDynamicProgramming2005}
Garud~N. Iyengar.
\newblock Robust {{Dynamic Programming}}.
\newblock \emph{Math. Oper. Res.}, 30\penalty0 (2):\penalty0 257--280, 2005.

\bibitem[Junges et~al.(2019)Junges, {\'A}brah{\'a}m, Hensel, Jansen, Katoen,
  Quatmann, and Volk]{jungesParameterSynthesisMarkov2019a}
Sebastian Junges, Erika {\'A}brah{\'a}m, Christian Hensel, Nils Jansen,
  Joost-Pieter Katoen, Tim Quatmann, and Matthias Volk.
\newblock Parameter {{Synthesis}} for {{Markov Models}}.
\newblock \emph{CoRR}, abs/1903.07993, 2019.

\bibitem[Kiwiel(2001)]{kiwielConvergenceEfficiencySubgradient2001}
Krzysztof~C. Kiwiel.
\newblock Convergence and efficiency of subgradient methods for quasiconvex
  minimization.
\newblock \emph{Math. Program.}, 90\penalty0 (1):\penalty0 1--25, 2001.

\bibitem[Knight(2002)]{knightSafetyCriticalSystems2002}
John~C. Knight.
\newblock Safety critical systems: Challenges and directions.
\newblock In Will Tracz, Michal Young, and Jeff Magee, editors,
  \emph{Proceedings of the 24th {{International Conference}} on {{Software
  Engineering}}, {{ICSE}} 2002, 19-25 {{May}} 2002, {{Orlando}}, {{Florida}},
  {{USA}}}, pages 547--550. {ACM}, 2002.
\newblock \doi{10.1145/581339.581406}.

\bibitem[Kozine and Utkin(2002)]{kozineIntervalValuedFiniteMarkov2002}
Igor Kozine and Lev~V. Utkin.
\newblock Interval-{{Valued Finite Markov Chains}}.
\newblock \emph{Reliab. Comput.}, 8\penalty0 (2):\penalty0 97--113, 2002.

\bibitem[Kwiatkowska et~al.(2011)Kwiatkowska, Norman, and
  Parker]{kwiatkowskaPRISMVerificationProbabilistic2011}
Marta~Z. Kwiatkowska, Gethin Norman, and David Parker.
\newblock {PRISM} 4.0: Verification of probabilistic real-time systems.
\newblock In \emph{{CAV}}, volume 6806 of \emph{Lecture Notes in Computer
  Science}, pages 585--591. Springer, 2011.

\bibitem[Meedeniya et~al.(2014)Meedeniya, Moser, Aleti, and
  Grunske]{meedeniyaEvaluatingProbabilisticModels2014}
Indika Meedeniya, Irene Moser, Aldeida Aleti, and Lars Grunske.
\newblock Evaluating probabilistic models with uncertain model parameters.
\newblock \emph{Softw. Syst. Model.}, 13\penalty0 (4):\penalty0 1395--1415,
  2014.

\bibitem[Nash(1989)]{nashNoncooperativeGames1989}
John Nash.
\newblock Non-cooperative games.
\newblock \emph{Cournot Oligopoly}, pages 82--94, January 1989.
\newblock \doi{10.1017/CBO9780511528231.007}.

\bibitem[Nilim and Ghaoui(2005)]{nilimRobustControlMarkov2005}
Arnab Nilim and Laurent~El Ghaoui.
\newblock Robust {{Control}} of {{Markov Decision Processes}} with {{Uncertain
  Transition Matrices}}.
\newblock \emph{Oper. Res.}, 53\penalty0 (5):\penalty0 780--798, 2005.

\bibitem[Osband et~al.(2019)Osband, Roy, Russo, and
  Wen]{DBLP:journals/jmlr/OsbandRRW19}
Ian Osband, Benjamin~Van Roy, Daniel~J. Russo, and Zheng Wen.
\newblock Deep exploration via randomized value functions.
\newblock \emph{J. Mach. Learn. Res.}, 20:\penalty0 124:1--124:62, 2019.

\bibitem[Platzer(2012)]{platzerLogicsDynamicalSystems2012}
Andr{\'e} Platzer.
\newblock Logics of {{Dynamical Systems}}.
\newblock In \emph{Proceedings of the 27th {{Annual IEEE Symposium}} on
  {{Logic}} in {{Computer Science}}, {{LICS}} 2012, {{Dubrovnik}}, {{Croatia}},
  {{June}} 25-28, 2012}, pages 13--24. {IEEE Computer Society}, 2012.
\newblock \doi{10.1109/LICS.2012.13}.

\bibitem[Porter et~al.(2008)Porter, Nudelman, and
  Shoham]{porterSimpleSearchMethods2008}
Ryan Porter, Eugene Nudelman, and Yoav Shoham.
\newblock Simple search methods for finding a {{Nash}} equilibrium.
\newblock \emph{Games Econ. Behav.}, 63\penalty0 (2):\penalty0 642--662, 2008.

\bibitem[Puggelli et~al.(2013)Puggelli, Li, {Sangiovanni-Vincentelli}, and
  Seshia]{puggelliPolynomialTimeVerificationPCTL2013a}
Alberto Puggelli, Wenchao Li, Alberto~L. {Sangiovanni-Vincentelli}, and
  Sanjit~A. Seshia.
\newblock Polynomial-{{Time Verification}} of {{PCTL Properties}} of {{MDPs}}
  with {{Convex Uncertainties}}.
\newblock In \emph{{{CAV}}}, volume 8044 of \emph{Lecture {{Notes}} in
  {{Computer Science}}}, pages 527--542. {Springer}, 2013.

\bibitem[Quatmann et~al.(2016)Quatmann, Dehnert, Jansen, Junges, and
  Katoen]{DBLP:conf/atva/QuatmannD0JK16}
Tim Quatmann, Christian Dehnert, Nils Jansen, Sebastian Junges, and
  Joost{-}Pieter Katoen.
\newblock Parameter synthesis for markov models: Faster than ever.
\newblock In \emph{{ATVA}}, volume 9938 of \emph{Lecture Notes in Computer
  Science}, pages 50--67, 2016.

\bibitem[Raskin and Sankur(2014)]{raskinMultipleEnvironmentMarkovDecision2014a}
Jean-Fran{\c c}ois Raskin and Ocan Sankur.
\newblock Multiple-{{Environment Markov Decision Processes}}.
\newblock In \emph{{{FSTTCS}}}, volume~29 of \emph{{{LIPIcs}}}, pages 531--543.
  {Schloss Dagstuhl - Leibniz-Zentrum f\"ur Informatik}, 2014.

\bibitem[Rickard et~al.(2023)Rickard, Badings, Romao, and
  Abate]{rickardFormalControllerSynthesis2023}
Luke Rickard, Thom Badings, Licio Romao, and Alessandro Abate.
\newblock Formal {{Controller Synthesis}} for {{Markov Jump Linear Systems}}
  with {{Uncertain Dynamics}}, May 2023.

\bibitem[Scheftelowitsch et~al.(2017)Scheftelowitsch, Buchholz, Hashemi, and
  Hermanns]{scheftelowitschMultiObjectiveApproachesMarkov2017}
Dimitri Scheftelowitsch, Peter Buchholz, Vahid Hashemi, and Holger Hermanns.
\newblock Multi-{{Objective Approaches}} to {{Markov Decision Processes}} with
  {{Uncertain Transition Parameters}}.
\newblock In \emph{{{VALUETOOLS}}}, pages 44--51. {ACM}, 2017.

\bibitem[Stackelberg(1952)]{stackelbergTheoryMarketEconomy1952}
H.~{\relax Von}. Stackelberg.
\newblock The {{Theory}} of {{Market Economy}}. {{Translated}} from the
  {{German}} and with an {{Introduction}} by {{A}}.{{T}}. {{Peacock}}.
  {{London}}, {{Edinburgh}}, {{Glasgow}}, {{W}}. {{Hodge}} \& {{Co}}, {{Ltd}}.,
  1952, xxiii p. 328 p., 25/-.
\newblock \emph{Recherches \'Economiques de Louvain/ Louvain Economic Review},
  18\penalty0 (5):\penalty0 543--543, 1952.
\newblock ISSN 1373-9719.
\newblock \doi{10.1017/S0770451800047382}.

\bibitem[van~der Vegt et~al.(2023)van~der Vegt, Jansen, and
  Junges]{vegtRobustAlmostSureReachability2023}
Marck van~der Vegt, Nils Jansen, and Sebastian Junges.
\newblock Robust {{Almost-Sure Reachability}} in {{Multi-Environment MDPs}}.
\newblock In \emph{{{TACAS}} (1)}, volume 13993 of \emph{Lecture {{Notes}} in
  {{Computer Science}}}, pages 508--526. {Springer}, 2023.

\bibitem[Wiesemann et~al.(2013)Wiesemann, Kuhn, and
  Rustem]{wiesemannRobustMarkovDecision2013}
Wolfram Wiesemann, Daniel Kuhn, and Ber{\c c} Rustem.
\newblock Robust {{Markov Decision Processes}}.
\newblock \emph{Math. Oper. Res.}, 38\penalty0 (1):\penalty0 153--183, 2013.

\bibitem[Yousefimanesh et~al.(2023)Yousefimanesh, Bos, and
  Vermeulen]{yousefimaneshStrategicRationingStackelberg2023}
Niloofar Yousefimanesh, Iwan Bos, and Dries Vermeulen.
\newblock Strategic rationing in {{Stackelberg}} games.
\newblock \emph{Games Econ. Behav.}, 140:\penalty0 529--555, 2023.

\bibitem[Yu and Dimarogonas(2021)]{yuDistributedMotionCoordination2021}
Pian Yu and Dimos~V. Dimarogonas.
\newblock Distributed motion coordination for multi-robot systems under {{LTL}}
  specifications, March 2021.

\end{thebibliography}

\ifappendix
\newpage
\appendix
\section{Proofs}
\label{app:proofs}

\subsection{Equivalence of memoryless behavioural and mixed policies}

\subsubsection{Realisation equivalent mixed policy for a given behavioural policy}
First, consider any behavioural policy $\pol^B \in \pols^B$.
We know that for an MDP, there will exist a deterministic policy with maximum probability of satisfying a PCTL formula, which we call $\overline{\pol^\star}$ (see \cite{baierPrinciplesModelChecking2008a} for a proof of this). 
More formally, we have that $$\overline{\pol^\star} \in \argmax_{\pol^B \in \pols^B}\max_{\lambda \in [0,1]} \PCTLp_{\geq \lambda} \pathForm,$$ recognising that a deterministic policy is a behavioural policy with all probability mass assigned to a single action in each state.

Hence, for an MDP induced by a given sample $v, \sol{\pol^B}{v} \leq \sol{\overline{\pol^\star}}{v}$.
If we consider trying to maximise for $\neg \pathForm$, it is obvious that there exists another optimal deterministic policy,  $\underline{\pol^\star}$, (since we are now optimising for a different, but equally valid PCTL formula).
 $$\underline{\pol^\star} \in \argmax_{\pol^B \in \pols^B}\max_{\lambda \in [0,1]} \PCTLp_{\geq \lambda} \neg\pathForm = \argmin_{\pol^B \in \pols^B}\min_{\lambda \in [0,1]} \PCTLp_{\leq \lambda} \pathForm,$$
 
This policy will then have a minimum probability of satisfying the original formula, $\pathForm$ so that $\sol{\pol^B}{v} \geq \sol{\underline{\pol^\star}}{v}$. 
Hence, we have $$\sol{\underline{\pol^\star}}{v} \leq \sol{\pol^B}{v} \leq \sol{\overline{\pol^\star}}{v}.$$
Finally, since a mixed policy defines a finite distribution over deterministic policies, it is straightforward to see that $$\sol{\pol^M}{v} = \sum_{\pol^D \in \pols^D}\pol^M(\pol^D)\cdot \sol{\pol^D}{v}.$$
Then, since $\sol{\pol^B}{v}$ is bounded from above and below there must exist a convex combination of these bounds that is equal to $\sol{\pol^B}{v}$, and this convex combination defines the realisation equivalent mixed policy.

\subsubsection{Realisation equivalent behavioural policy for a given mixed policy}
A similar argument can be made for finding a behavioural policy for a given mixed policy $\pol^M \in \pols^M$.
However, for behavioural policies it does not, in general, hold that a convex combination of policies will provide an equally weighted convex combination of values.
Specifically, for $\gamma \in (0,1)$, $\pol^B_1, \pol^B_2 \in \pols^B$:  $$\sol{\gamma\pol^B_1+(1-\gamma)\pol^B_2}{v} \neq \gamma \sol{\pol^B_1}{v} + (1-\gamma)\sol{\pol^B_2}{v}.$$
We note that there are limited cases where this may in fact hold with equality. 
For example, for very simple MDPs with a single state having more than one enabled action. 

Instead, we rely on the intermediate value theorem. 
As with the previous subsection, for all $\pol^B \in \pols^B$, there exist $\underline{\pol^\star}, \overline{\pol^\star}$, such that 
$$\sol{\pol^B}{v} \in [\sol{\underline{\pol^\star}}{v},\sol{\overline{\pol^\star}}{v}],$$
and for our given mixed policy, $\sol{\pol^M}{v} \in [\sol{\underline{\pol^\star}}{v},\sol{\overline{\pol^\star}}{v}].$
Hence, if $\sol{\pol^B}{v}$ can be shown to be continuous in $\pol^B$, then it must hold that $$\exists \pol^B \in \pols^B \colon \sol{\pol^B}{v} = \sol{\pol^M}{v}.$$

Since the discount factor is strictly less than 1, there will not be any discontinuities in the solution function\footnote{If this assumption is violated, discontinuities may arise in the presence of PCTL formulae with infinite horizons where loops which can be followed with certainty exist.
For a concrete example, consider a simple MDP with an initial state and a goal state, and two actions in the initial state corresponding to transitioning to the goal state or staying in the initial state (both with probability 1), with a discount factor equal to 1. 
The PCTL formula is simply an infinite horizon reach probability. 
For any policy taking the action to transition to the goal state with a non-zero probability, the formula will be satisfied with probability 1. 
However, for the single policy which always transitions back to the initial state (i.e. taking the action to transition with probability 0), the reach probability will also be 0.
Hence, a discontinuity arises.
If the loop probability is less than 1, then this discontinuity is removed. 
This example clearly also violates our equivalence.}

To see this, consider $\sol{\pol^B}{v}$ encoding a PCTL formula $\stateForm = \PCTLp_{\geq \lambda}\text{S} \until \text{G}$, with state $s_G$ satisfying G, this may be expanded as 
\begin{align*}
&\sol{\pol^B}{v} =\\ &\sum_{s \in \states} \rho(s)\sum_{a \in \acts(s)} \pol^B(a \given s) \cdot \Bigg(\gamma \probs(s_G \given s,a)+ \\
&\sum_{s' \in \states \setminus s_G} \gamma \probs(s' \given s, a) \left[  \sum_{a' \in \acts(s')} \pol^B(a' \given s') \cdot \big( \gamma \probs(s_G \given s',a')+ \sum_{s'' \in \states\setminus s_G} \gamma \probs(s'' \given s', a') \dots \big)\right]\Bigg). 
\end{align*}
Since this is a summation and multiplication of many terms linear in the policy, and all infinite series involved are convergent, it is continuous in the policy. 
Thus, the intermediate value theorem will hold, and for any mixed policy we there exists an associated behavioural policy which is realisation equivalent (although the statement is not constructive).
Further PCTL formulae are explored in \cref{app:PCTL_Q}.

\section{Details on PCTL Verification}
\label{app:PCTL_Q}

\subsection{Defining a Q-Function}

For any general PCTL formula, we can define an associated Q-function as follows.
First, note that all operators in PCTL may be derived from the until operator \cite{hanssonLogicReasoningTime1994}, as such we consider only formulae with this operator.
Second, we will at first only consider non-nested formulae, and look into nested formulae in \cref{app:PCTL_Q:nest}.

Then, our formula is of the form $$\stateForm = \PCTLp_{\geq \lambda}\text{S} \until \text{G},$$ and we consider states $\states_S,\states_G \subseteq \states$ to refer to the states labelled with atomic propositions S and G respectively.

For a given sampled MDP $\sampledMDP$, we can then define an iterative bellman equation $\mathcal{B}_{\pol_B}^v \colon \states \rightarrow [0,1]$ as follows (we consider only behavioural policies here, since the other classes of policy can be solved without the use of a Q-function):
\begin{equation}
    \mathcal{B}_{\pol_B}^v(s) \coloneqq \begin{cases}
    1 & \mathrm{if } \; s \in \states_G,\\
    0 & \mathrm{if } \; s \notin \states_G \; \mathrm{and} \; s \in \states \setminus \states_S,\\
    \sum_{a \in \acts} \pol_B(s)(a) \sum_{s' \in \states} \paramProbs_v(s,a)(s') \mathcal{B}_{\pol_B}^v(s') & \mathrm{otherwise}.
    \end{cases}
\end{equation}
This naturally models the probability of a given state $s$ satisfying the PCTL path formula $\text{S} \until \text{G}$.

Finally, define the Q-function using this bellman equation as
\begin{equation}
    Q_{\pol_B}^v(s,a) \coloneqq \sum_{s' \in \states} \probs_v(s,a)(s') \mathcal{B}_{\pol_B}^v(s').
\end{equation}

\subsection{Nested Formulae}
\label{app:PCTL_Q:nest}

For nested formulae, we simply consider using a computation tree (as is done in \cite{rickardFormalControllerSynthesis2023}).
Thus, we break a nested formula down until the leaves represent simple reach-avoid formulae $\PCTLp_{\geq \lambda}\text{S} \until \text{G}$, we optimise the probability of satisfying these formulae, and find states exceeding (or below, depending on the direction of the inequality) the specified bound $\lambda$.
States that satisfy the bound can be passed up the tree as goal states (if appearing to the right of an until operator), or as safe states (if on the left of an until operator).
\section{Experimental Details}
\label{app:exp_det}

\subsection{UAV Motion Planning}

We use a model very similar to the UAV motion planning problem introduced in \cite{badingsScenariobasedVerificationUncertain2022} (with the addition of a discount factor).
This model consists of a grid world in which the UAV can decide to fly in either of the six cardinal directions (N, W, S, E, up, down).
States encode the position of the UAV, the current weather condition (sunny, stormy), and the general wind direction in the valley. 
The probabilistic outcomes are assumed to be determined only by the wind in the valley and the control action. 
An action moves the UAV one cell in the corresponding direction, and additionally, the wind moves the UAV one cell in the wind direction with a probability $p$, (dependendt on the wind speed). 
We assume that the weather and wind-conditions change during the day and are described by a stochastic process.
Concretely, parameters describe how the weather affects the UAV in different zones of the valley, and how the weather/wind may change during the day. 

We make 3 different assumptions on the distribution over the parameters, and call these ``uniform'', ``x-neg-bias'' and ``y-pos-bias'', they are defined as follows:

\paragraph{uniform} a uniform distribution over the different weather conditions in each zone;
\paragraph{x-neg-bias} the probability for a weather condition inducing a wind direction that pushes the UAV westbound (i.e., into the negative x-direction) is twice as likely as in other directions;
\paragraph{y-pos-bias} it is twice as likely to push the UAV northbound (i.e., into the positive y-direction).

\subsection{Benchmark Models}

The benchmark models (``consensus'', ``brp'', ``sav'', ``zeroconf'') are taken from \cite{DBLP:conf/atva/QuatmannD0JK16}. 

\subsection{Toy Model}
In order to compare all algorithms presented, we consider a very small toy model.
For this model, each state has two actions, the first which may take us to the next state or the critical state, and the second which may jump us forward two states, or take us to the critical state. 
We consider there being two non-absorbing states.
Then, the probability of an action being successful is a parameter for each state action pair, with the remaining probability being assigned to us ending up in the critical state.
Our goal is simply to reach the final state without visiting the critical state.

\subsection{Model Sizes}

Further details on model sizes are presented in \cref{tab:model_sizes} (where bisimulations can be used to reduce the model size we consider the reduced model).
Recall that the MNE algorithm requires iterating through all deterministic policies (on the order of $|\acts|^{|\states|}$), hence it should be clear why this algorithm scales so poorly, with the smallest non-toy model having already on the order of $10^{46}$ policies.

\begin{table}[h]
    \centering
    \begin{tabular}{c | c c c c}
        \hline
          & \#Pars & \#States & \#Actions & \#Transitions \\ \hline \hline 
          \textbf{consensus} & & & & \\
          (2,2) & 2 & 153 & 2 & 332  \\
          (2,32) & 2 & 2 793 & 2 & 6 092 \\
          (4,2) & 4 & 22 656 & 4 & 75 232 \\
         \hline
        \textbf{brp}   \\
        (256,5) & 2 & 21 032 & 2 & 29 750  \\
          (4096,5) & 2 & 647 441  & 2 & 876 903 \\
         \hline 
       \textbf{sav}\\
        (6,2,2) & 2 & 379 & 4 & 1 127  \\
         (100,10,10) & 2 & 1 307 395 & 4 & 6 474 535\\
          (6,2,2) & 4 & 379 & 4 & 1 127\\
         (10,3,3) & 4 & 1 850 & 4 & 6 561 \\
         \hline
         \textbf{zeroconf}  \\
        (2) & 2 &  88 858 & 4&203 550 \\
         (5) & 2 & 494 930 & 4& 1 133 781\\
         \hline
         \textbf{UAV}  \\
         uniform & 900 & 7642 & 6  & 56 568 \\
         x-neg-bias & 900& 7642 & 6  & 56 568 \\
         y-pos-bias &  900 & 7642 & 6  & 56 568\\
         \hline
         \textbf{Toy Model} & 4 & 4 & 2 & 10 \\
         
         \hline
         \end{tabular}
    \caption{Model Sizes}
    \label{tab:model_sizes}
\end{table}


\section{Additional Results}
\label{app:add_res}

\subsection{Subgradient Algorithm Behaviour}

We provide two plots here which demonstrate convergent behaviour of our proposed subgradient algorithm.
The first, in \cref{fig:test_dist}, is for a small test model where we can use the MNE algorithm to find a true optimal mixed policy (and hence a true optimal satisfaction probability).
Here we can see that the distance to this optimal decreases across iterations, and we get remarkably close to the true optimal value.
Note that we do not have a formal proof that a behavioural policy can reach this theoretical optimum, so this quite a remarkable result since it demonstrates that our algorithm does appear to converge to the true optimum.

Second, in \cref{fig:UAV_dist}, we consider the UAV benchmark from \cref{tab:exp_res}, with uniform wind conditions, and compare the distance to the final satisfaction probability.
In this case, the probability once again appears to show convergent behaviour and the distance is continually decreasing.

\begin{figure}
    \centering
    \includegraphics[width=.7\textwidth]{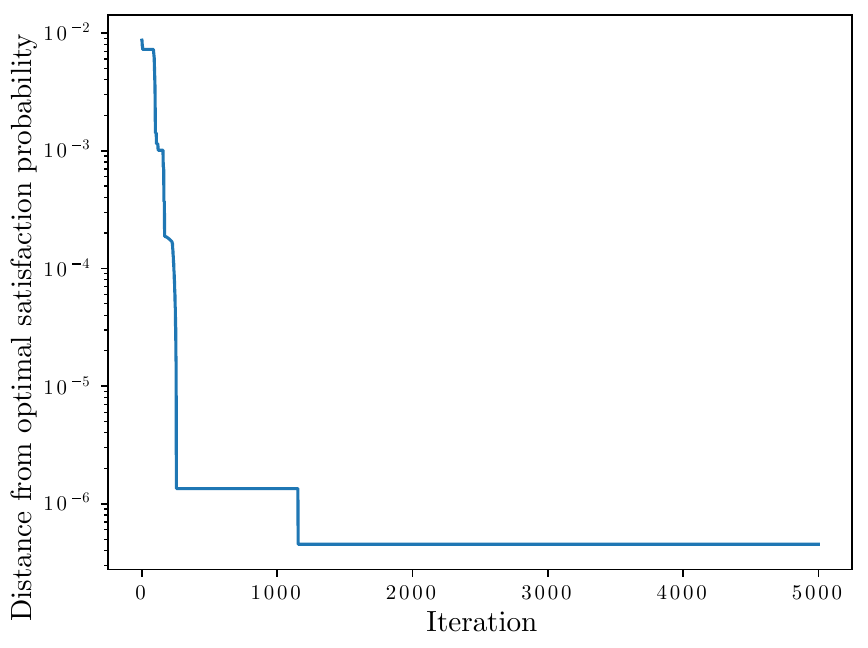}
    \captionof{figure}{Distance from optimal satisfaction probability (found from MNE algorithm) across iterations for small test model}
    \label{fig:test_dist}
\end{figure}

\begin{figure}
    \centering
    \includegraphics[width=.7\textwidth]{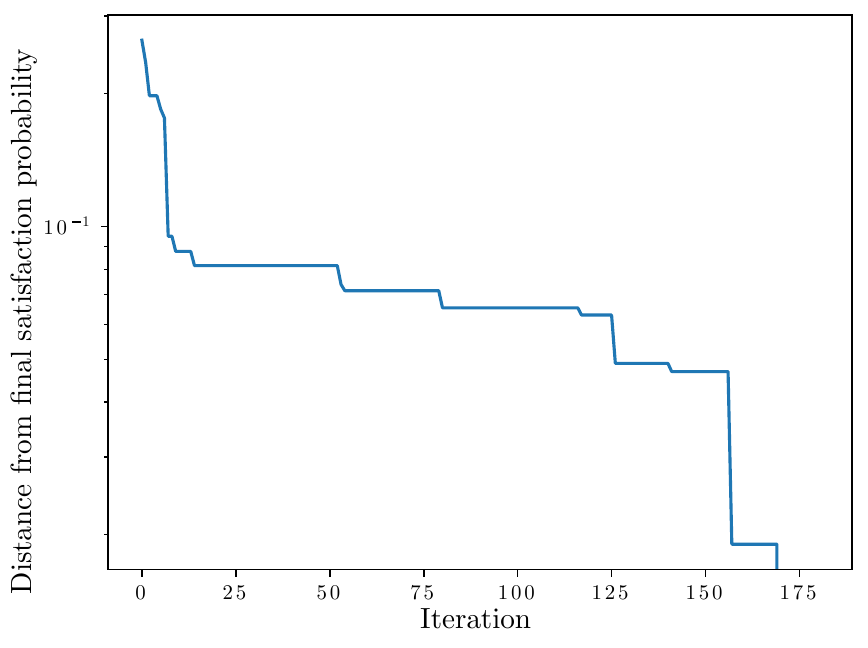}
    \caption{Distance from final satisfaction probability across iterations for UAV model with uniform wind}
    \label{fig:UAV_dist}
\end{figure}

\subsection{Runtimes}

We have also considered how the various methods presented scale with both the number of sampled parameters, and the size of the MDP.
In \cref{fig:sample_time}, we see how the number of samples $N$ affects the runtime, evidently the work of \cite{badingsScenariobasedVerificationUncertain2022} scales approximately linearly (since they solve each sampled MDP).
The deterministic solver based on the MaxMin game (explored further in \cref{app:det}) and an MNE solver making use of fictitious self play to approximately solve for the equilibrium, clearly scale linearly, since they both must iterate through all samples to construct a reward matrix, but further computation is largely negligible.
The solution based on iMDPs appears almost constant with respect to sample size, since we only need to find the maximum and minimum probabilities from amongst the samples this is roughly as expected (in fact we should expect linear scaling, but solving the iMDP takes longer and is independent of the sample size).
Finally, for the subgradient and MNE solver using a PNS algorithm the results are more difficult to interpret, but may also be linear, as may be expected.

\begin{figure}
    \centering
    \includegraphics[width=.7\textwidth]{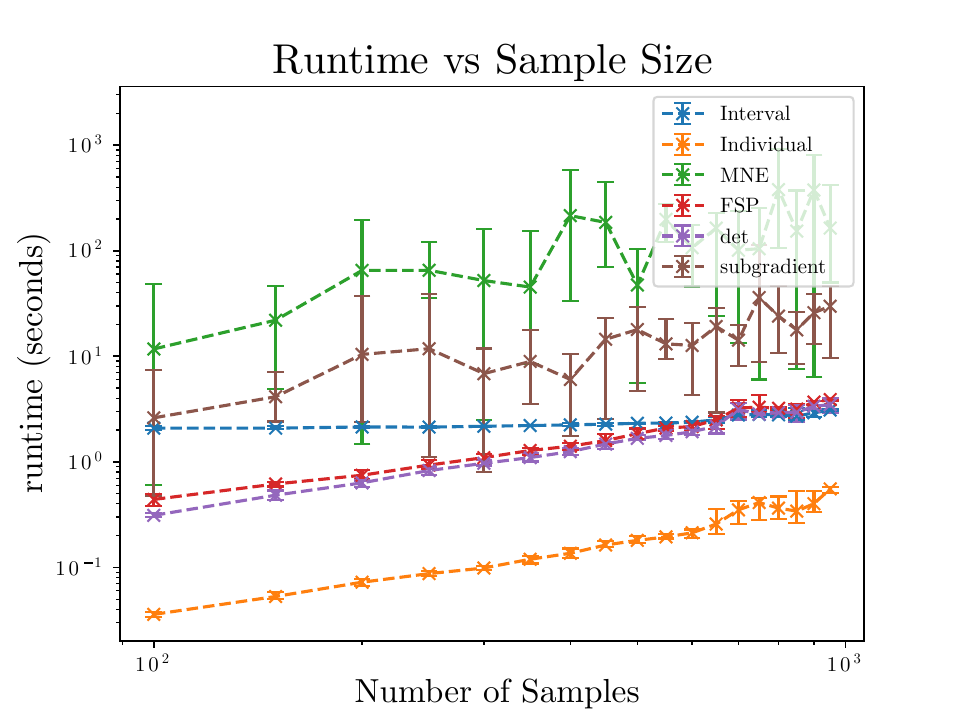}
    \caption{Runtime against number of samples}
    \label{fig:sample_time}
\end{figure}

In \cref{fig:size_time}, we consider how the size of the MDP affects the runtimes of our various algorithms.
The algorithms which require calculating a reward matrix scale approximately exponentially in the MDP size, and quickly start reaching computation limits.
The other algorithms (ignoring some spurious initial behaviour), seem to scale roughly linearly with MDP size. 
For our subgradient algorithm this is as expected, since for each iteration we update every state-action pair individually.
For the other algorithms, this is likely related to the model checker used to solve the models.

\begin{figure}
    \centering
    \includegraphics[width=.7\textwidth]{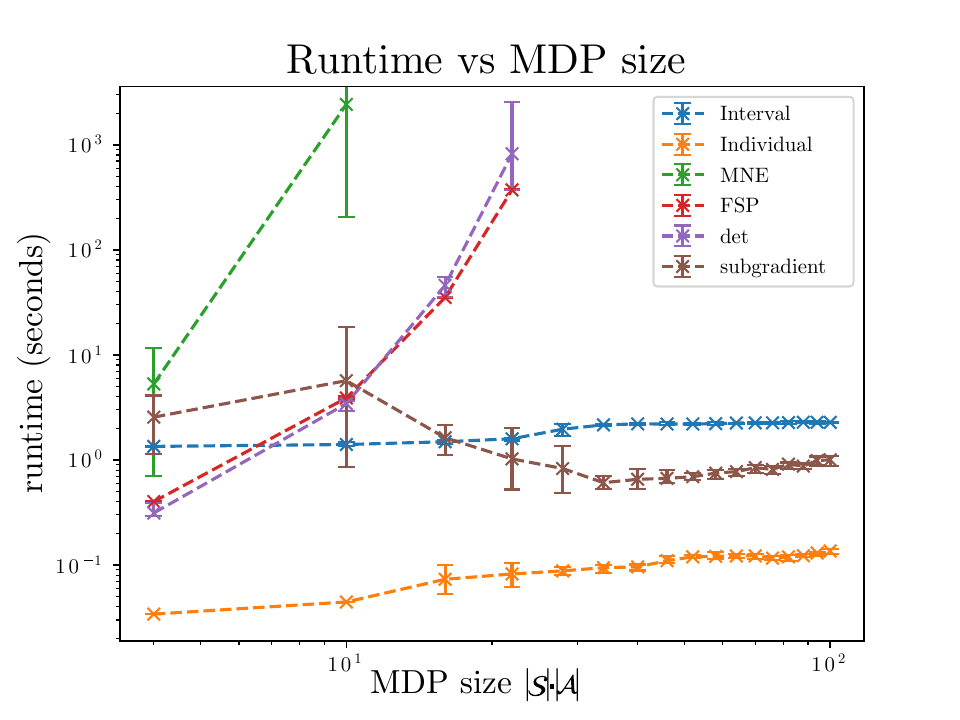}
    \caption{Runtime against size of MDP}
    \label{fig:size_time}
\end{figure}
\section{Deterministic MaxMin Policies}
\label{app:det}

\subsection{Finding MaxMin Deterministic Policies}

One solution concept of interest involves finding the maximin deterministic policies for the game introduced in \cref{sec:robust:Mixed}, through the use of a Stackelberg game \citep{stackelbergTheoryMarketEconomy1952}. 
To achieve this, the sample adversary acts as the leader and finds a minimising sample for \emph{each} possible deterministic policy, that is, for every policy $\pi^D$, the adversary solves
\begin{equation}
    r_p(\pol^D) = \min_{u \in \allSamples} \sol{\pol^D}{u},
\end{equation}
then the policy agent acts as the follower and chooses a deterministic policy which maximises the reward $r_p(\pol^D)$.
In this case, we are working with finite sets of actions for both agents, and so we can consider simply enumerating over all choices.

The resulting policy is the optimal deterministic policy, but may not be a Nash equilibrium policy (see \cref{thm:NE}). 
To see this note that if the sample adversary takes a fixed strategy (as opposed to always being allowed to choose the worst case), the policy agent may be able to improve their reward.

\subsection{Probabilistic Guarantees under a MaxMin Deterministic Policy}

In this case, our problem is degenerate, since it may be necessary to remove multiple constraints in order for another deterministic policy to become optimal, we therefore leverage techniques from \cite{campiGeneralScenarioTheory2018} for non-convex problems. 
This approach requires solving a different equation for the risk, $\mu(\supps)$, which must satisfy \cite[Theorem 1]{campiGeneralScenarioTheory2018}.
One equation which satisfies these requirements is that in \cref{eq:det_risk}.

The easiest method for obtaining an upper bound on the size of the support set is to consider the satisfaction probabilities for each deterministic policy in the worst-case MDP $\optSampledMDP$ for the optimal policy.
If the satisfaction probability for a different policy is greater than the worst-case, then there must exist at least one sample that prevents us switching to that policy.
As an upper bound, we may consider that every such policy is ``blocked'' by a single unique sample:

Alternatively, to improve this bound, we first find the set of ``blocked'' policies as above.
Then, for each such policy, we find the samples which have an associated satisfaction probability less than our optimal $\lambda^\star$.
To calculate the smallest support set, we then must find the smallest set such that at least one blocking sample for every policy is contained within this set. 
This problem is known as the hitting set problem and can be shown to be NP-hard.
Alternatively, we may take a union of all sets to again find an upper bound.

Note that both methods may be computationally expensive if there are many possible deterministic policies.

We then have the following result that bounds the risk:

\begin{corollary}
    We use \cref{eq:det_risk} to bound the risk.
    Bounds on the support constraints may be found as
    \begin{equation}
        \overline{\supps} = |\{\pol^D \in \pols^D \colon Pr(\pathForm \given \pol^D, \optSampledMDP) \geq \lambda^\star\}|,
    \end{equation}
    or 
    \begin{align}
            \AllSuppSet(\pol^D) &= \{ u \in \allSamples \colon \sol{\pol^D}{u} \leq \lambda^\star \wedge  \sol{\pol^D}{u^\star} \geq \lambda^\star \}\\
            \supps &= \min_{\mathcal{V} \subseteq \allSamples} |\mathcal{V}| \; \text{subject to:}\; \mathcal{V} \bigcap \AllSuppSet(\pol^D) \neq \emptyset, \forall \pol^D  \in \pols^D, \label{eq:hitting}
    \end{align}
    the latter being less conservative, but significantly more computationally expensive.
    Approximate solutions to \cref{eq:hitting} exist which may allow for a tradeoff between computation and optimality.

    Then we have that
\begin{align}
        \mathbb{P}^N\left\{\mathbb{P}\left\{v \in \params \colon \sampledMDP \notsat_{\pi^D} \PCTLp_{\geq \lambda^\star}(\pathForm) \right\} \leq \mu(\supps) \leq \mu(\overline{\supps})\right\} \geq 1-\beta,
\end{align}
where the only non-determinism arises due to the uncertainty in the transition function.
\end{corollary}

\subsection{Results for Deterministic Policies}

As is the case for our MNE algorithm, finding deterministic policies requires iterating through all possible deterministic policies.
As such, we only have experimental results for our small toy model, we recall the results for our algorithms on this model, and provide results for the deterministic policy in \cref{tab:det_res}.

\begin{table}[]
    \centering
    \begin{tabular}{c|c c c c c}
         & Sat. Prob. $\lambda^\star$ & Risk U.B. $\overline{\epsilon}$ & Time (s) & Emp. Risk $\Tilde{\epsilon}$ & Emp. (no pars) \\ \hline
         \cite{badingsScenariobasedVerificationUncertain2022} & .338 & .056 & .07 & .018 & .115\\
         iMDP Solver & .329 & .155 & 1.39 & .015 & .107 \\
         MNE Algorithm & .333 & .109 & 4.51 & .043 & .121 \\
         Subgradient Algorithm & .332 & .285 & 26.12 & .022 & .115 \\
         Deterministic MaxMin Policy & .329 & .105 & 0.28 & .012 & .105 
    \end{tabular}
    \caption{Results on Toy Model, with Deterministic Policies}
    \label{tab:det_res}
\end{table}
\fi
\end{document}